\documentclass{article}

\usepackage[utf8]{inputenc}
\usepackage[T1]{fontenc}

\usepackage[table]{xcolor}
\usepackage{graphicx}
\usepackage{multirow}
\usepackage[normalem]{ulem}
\usepackage{subcaption}
\usepackage{wrapfig}
\usepackage{amsthm} % For theorems

\usepackage{tikz}
\usepackage{pgfplots}
\usepackage{pgfplotstable}
\usepackage{accessibility}
\usepackage{amsmath}
\usepackage{hyperref}

\usepackage{authblk}

\pgfplotsset{compat=1.7}
\usetikzlibrary{arrows.meta, patterns}

\newcommand{\blue}[1]{\textcolor{black}{#1}}

\tikzset{
  bar/.tip={Rectangle[length=1pt, width=3pt]},
  signbarL/.style={
    {bar[right]}-{bar[left]}
  }
}

\pgfplotsset{
    width=7cm,
    /pgfplots/ybar legend/.style={
        /pgfplots/legend image code/.code={
           \draw[##1,/tikz/.cd,yshift=-0.25em]
            (0cm,0cm) rectangle (3pt,0.8em);
        },
    },
}

\title{Cooperation Through Indirect Reciprocity in Child-Robot Interactions}

\author[1]{Isabel Neto\thanks{Equal contribution}}
\author[2]{Alexandre S. Pires\textsuperscript{*}}
\author[3]{Filipa Correia}
\author[2]{Fernando P. Santos}

\affil[1]{Lasige, Faculdade de Ciências, Universidade de Lisboa, Lisbon, Portugal}
\affil[2]{Institute of Informatics, University of Amsterdam, Amsterdam, The Netherlands}
\affil[3]{Interactive Technologies Institute, Instituto Superior Técnico, Universidade de Lisboa, Lisbon, Portugal}

\begin{document}

\maketitle

%%%% Abstract text to be placed here %%%%%%%%%%%%
\begin{abstract}
Social interactions increasingly involve artificial agents, such as conversational or collaborative bots. Understanding trust and prosociality in these settings is fundamental to improve human-AI teamwork. Research in biology and social sciences has identified mechanisms to sustain cooperation among humans. Indirect reciprocity (IR) is one of them. With IR, helping someone can enhance an individual’s reputation, nudging others to reciprocate in the future. Transposing IR to human-AI interactions is however challenging, as differences in human demographics, moral judgements, and agents’ learning dynamics can affect how interactions are assessed. To study IR in human-AI groups, we combine laboratory experiments and theoretical modelling. We investigate whether 1) indirect reciprocity can be transposed to children-robot interactions; 2) artificial agents can learn to cooperate given children’s strategies; and 3) how differences in learning algorithms impact human-AI cooperation. We find that IR extends to children and robots solving coordination dilemmas. Furthermore, we observe that the strategies revealed by children provide a sufficient signal for multi-armed bandit algorithms to learn cooperative actions. Beyond the experimental scenarios, we observe that cooperating through multi-armed bandit algorithms is highly dependent on the strategies revealed by humans. 
\end{abstract}
%%%%%%%%%%%%%%%%%%%%%%%%%%%

\maketitle

\section{Introduction}\label{sec1}

% Introduction with related work 
% Isabel 

% problem? with the increase use or presence of artificial agents / robots in our society, how robot's reputation influences cooperation in hybrid groups ? And how it evolve to time 

Social robots are increasingly part of humans' daily routines, including in homes and schools \cite{cagitay2024, belpaeme2018social}. As a result, children are exposed to novel interaction paradigms with artificial agents, both individually and in groups. In such domains, it is fundamental to understand how artificial agents might impact children's social dynamics, particularly their willingness to cooperate.  It is also desirable to design robots that can engage and cooperate effectively within these hybrid groups \cite{akata_research_2020, jennings2014human, dafoe2021cooperative}. Ensuring cooperation and fostering lasting interactions in such environments is crucial to enable a positive integration of social robots into children's daily lives. 
In the near future, it is expected that robots will serve as tutors, companions, and \blue{educational} tools \cite{belpaeme2018social}, as well as delivery and security agents in public spaces \cite{pelikan2025making} or assistants in our homes \cite{cagitay2024}. As these interactions become more common, several challenges are emerging, and effective robot cooperation strategies remain unexplored. For example, how can robots collaborate without overshadowing children’s intentions in physically demanding activities such as games and sports? How can they act as tutors without completing parts of schoolwork on children's behalf, thus reducing learning opportunities? Or even, how can robots adapt their behaviour to balance family participation in group activities while fostering a cooperative and inclusive environment?

% motivation  Indirect reciprocity is a mechanism of cooperation in humans population. Even in children.

Cooperation mechanisms, previously studied in biology \cite{nowak2006five} or \blue{psychology \cite{margoni2025rise}}, could help facilitate smoother human-AI interactions \cite{rand2013human}. An example of cooperation mechanisms is indirect reciprocity (IR) \cite{alexander2017biology}, which has been extensively studied in human populations \cite{nowak_evolution_2005, van2015economics, young2015evolution}. IR suggests that an individual's behaviour towards others is influenced not only by direct interactions, but also by the reputation received from third parties \cite{okada_review_2020,schmid2021unified}. 

The relevance of reciprocity and reputations in human social interactions is emphasized by the early age at which we use this information to manage interpersonal relations. Even 2-year-old children attribute reputations to others based on their past actions \cite{Hamlin2011, jin2024infants, nava2019see, meristo2013infants}. Not only do children recognise others' reputations, but they also use this information to guide their behaviour through indirect reciprocity. They tend to respond with positive and cooperative behaviours towards prosocial cooperative individuals while exhibiting negative and non-cooperative behaviours towards selfish non-cooperative individuals \cite{Hamlin2011, kato2013preschool, nava2019see, holvoet2016infants, margoni2025rise}. Notably, this reputation-based mechanism is not only limited to children interacting with each other but also extends to how children interact with artificial virtual agents \cite{meristo2013infants}.

% existing work on adults , IR and agents 
% Alex - prior works in IR modelling; multi-armed bandits for cooperation;
Several recent works have focused on \blue{the influence of artificial agents on long-term prosocial and cooperative behaviours of adults --- for example, through robot empathy, ingroup favouritism, agent delegation,} and different payoffs for humans and machines \cite{fernandez2022delegation,  paiva2021empathy,  romano2024nasty, santos2020picky, vonshenk25Systematic}. \blue{Theoretical models have also been developed to assess the impact of artificial agents employing fixed strategies in human cooperation \cite{pires2024artificial}.} It remains unclear, however, whether these findings and models apply to children\blue{, and whether agents can learn to respond using cooperative strategies}.
% apply to children
To our knowledge, only one study has investigated reciprocity in children-robot interactions, particularly robotic ostracism with children \cite{correia24ostracism}. Research in this area is limited, and the long-term effects of such interactions within hybrid groups remain largely unexplored.  

% research questions 
This paper explores three key research questions:
Can indirect reciprocity be transposed to Child-Robot Interactions? (RQ1); Can a robot learn to cooperate given the strategies revealed by children? (RQ2); What is the impact of different robotic behaviours and children's strategies on cooperation? (RQ3)
% approach : use empirical studies to understand children behaviour and model their long-term interactions

To address these research questions, we first conducted an empirical study with 131 participants, adapting the Stag Hunt game \cite{skyrms2001stag} for hybrid groups of children and robots to play with physical LEGO bricks. Our objective was to examine whether a child would \blue{cooperate (i.e., trust)} or \blue{defect (i.e., not trust)} when playing with a robot that had previously played the same game in either a cooperative or non-cooperative way with another child. 
% game details 
In this experiment, the participant \blue{was informed that her game rewards would contribute to a team goal. She} first observed a game interaction between a partner child and the robot. While the partner was privately instructed beforehand to always cooperate, the robot was programmed to either cooperate or defect according to the experimental condition (i.e. cooperative vs. non-cooperative).
% experimental conditions 
%In the \textbf{cooperative condition}, both the first player and the robot played their three bricks, and each gained two additional bricks, as an extra benefit ($b=2$). In the \textbf{non-cooperative condition}, the first player played their three bricks while the robot kept its three bricks for itself, causing the first player to lose its bricks. 
%
\begin{figure*}[h!]\centering
\includegraphics[width=\textwidth]{images/Overview_Compact_Model_recolor.png}
\caption{\textbf{A} Experimental setup: Children and robot play a social dilemma where cooperation is socially beneficial yet risky. If both individuals cooperate, they both receive 2 extra bricks (here used as a unit of payoff), however unilaterally cooperating leads to 0 payoff to the cooperator. The 2 additional bricks received are here referred to as the extra benefit of mutual cooperation ($b=2$). \textbf{B} The chain interaction model stemming from the  setup presented in Figure \ref{fig:overview_experiment}\textbf{A}, and motivating our simulation experiments. The artificial agent repeatedly plays against an individual, and is observed by another player. After the interaction, the artificial agent plays with the individual previously acting as observer and a new observer joins.}
\label{fig:overview_experiment}
\end{figure*}

After this observation phase, the participant played the same game with the robot, independently deciding whether to cooperate or not. 
\blue{Finally, the child contributed her bricks to a collective goal of building a brick tower. An illustration of the experimental design is presented in Figure \ref{fig:overview_experiment}A.}
%The second player’s cooperation rate served as the key independent variable in our study.
% output metrics 
The cooperation rate, defined as the proportion of children who chose to play with their bricks (i.e., to cooperate) in the game relative to the total number of children per condition --- $\hat{p}$ for cooperative condition and $\hat{q}$ for non-cooperative condition ---, served as the key dependent variable in our study and was then compared between experimental conditions. %then analyzed to inform the development of evolutionary models and mapped to $p$ and $q$. These models allowed us to explore the long-term effects of indirect reciprocity on child-robot cooperation. An illustration of the experimental setup and the theoretical model is presented in Figure \ref{fig:overview_experiment}.
% highlight results 
% reciprocidade indirecta aplica-se em grupos hibridos (fig 2)
The results indicate that children attribute a reputation to robots based on their observed cooperative (trust) or non-cooperative (defect) behaviour toward another child. Furthermore, this perceived reputation directly influences how children interact and cooperate with the robot when they engage with it in play afterward. These findings demonstrate that indirect reciprocity, as a mechanism of cooperation and trust, can be extended to hybrid groups, where children adjust their behaviour according to previous actions of a robot, see Figure \ref{fig:cooperative}.

% no imediato, o robô pode aprender a cooperar mas tem de se usar o algoritmo certo  (fig 3) 
Using the cooperation rate from our experimental results -- $\hat{p}$ for cooperative condition and $\hat{q}$ for non-cooperative condition -- , we built a theoretical multi-armed bandit (MAB) \cite{slivkins2019introduction} model to assess how different learning algorithms develop cooperative behaviour. In our model, the algorithm plays a repeated version of the Stag Hunt conducted in the experiment, where in each round it faces the observer of the prior interaction, allowing us to understand its long-term behaviour. Importantly, although the model defines a non-stationary problem \cite{besbes2014stochastic} -- as the payoff of each action is dependent on the prior action of the algorithm --, we opt to define the algorithms without assumptions about the behaviour of children. This allows us to explore the challenges of introducing a minimal algorithm while not depending on potentially non-Markovian formulations of children's strategies. This contrasts prior works which aim to maximise cooperation through complex stateful modelling \cite{crandall2018cooperating}. \blue{The chained interactions of the algorithm are presented in Figure \ref{fig:overview_experiment}B.} We study three different types of traditional MAB algorithms: $\epsilon$-greedy \cite{slivkins2019introduction}, UCB1 \cite{auer2002finite}, and Thompson sampling (TS) \cite{thompson1933likelihood}. In addition to simulating the experimental setup, we study the limits under which each of the different algorithms learns to cooperate, depending on the benefit obtained from mutual cooperation and on how frequently the algorithm receives cooperation after it cooperates ($p$) or defects ($q$). 

% a qualidade do algoritmo depende de contexto  , se no futuro o p e q alterar hipotetico e noutros cenários (fig 4) 
By studying how changes in the behaviour of children regarding cooperating with the robot affect the cooperation rate of the algorithm, we clarify under which contexts each algorithm is best suited for, and provide a baseline for other cultural environments and future scenarios. We observe that each algorithm reacts differently to either more frequent punishment following its defection, or more common cooperation after it cooperates.

% contributions 
This paper provides three key contributions. First, it presents empirical evidence of indirect reciprocity in Child-Robot Interaction. Specifically, we provide the first investigation into mechanisms of indirect reciprocity (IR) in hybrid groups of robots and children, and whether the observed robot cooperative (vs. non-cooperative) behaviour influences children's future cooperation. Our findings demonstrate that children's perception of a robot's reputation, based on prior interactions, shapes their subsequent cooperative behaviour.

Second, it bridges empirical studies and learning dynamics models. We introduce novel approaches for integrating empirical findings into simulation models to explore the long-term impact of robot design decisions. Specifically, our experiment informed the cooperation rate of children towards a robot to be used in the creation of a predictive model. Finally, we contribute with a thorough discussion of the implications of robotic behaviours and their design options for using different AI learning methods in the development of social robots that foster positive and cooperative Child-Robot Interactions.

% fig 1 . teaser :  experiencia  + modelo 

\section{Results }\label{sec2}

\subsection{Empirical Study}

Our empirical study considers a 2-stage coordination game, where in stage one a child observes a child-robot interaction, and in stage two the same child decides to cooperate or defect with the robot. The payoff structure of the underlying game assumes that mutual cooperation is payoff dominant (i.e., leads to the highest possible payoffs to all players), yet unilateral cooperation results in the lowest possible payoff to the cooperator. This interaction is close to the well-known Stag Hunt. In a nutshell, for this coordination dilemma to be solved and coordination failure avoided, humans need to trust the artificial agent. This setup allows us to analyse how children's trust in the robot in a cooperation dilemma is influenced by their perception of a robot's prior cooperative behaviour in an indirect reciprocity context.

In this study, we had 131 participants, all children aged 8 to 10 years ($M = 8.664$, $SD = 0.640$). Our pre-registered target sample size was 128 children \cite{preregister}. Participants were divided into two conditions: the cooperative condition (58 children observed the robot sharing the LEGO bricks and behaving cooperatively before playing with it) and the non-cooperative condition (58 children observed the robot behaving non-cooperatively, keeping the LEGO bricks, before playing with it).
Fifteen children were excluded from the analysis as the predefined procedure was not followed. Consequently, the final dataset for the statistical analysis included the play behaviour of 116 children, half in each condition.  

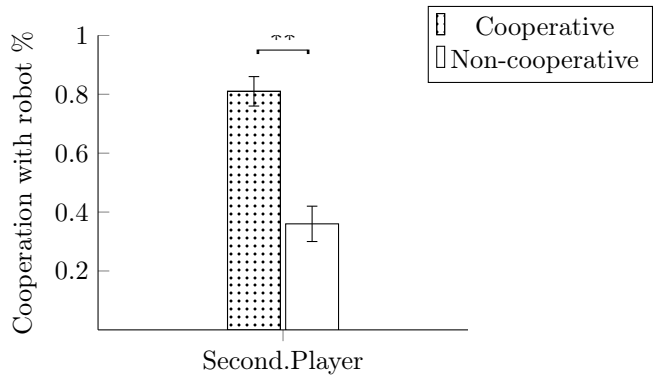
\begin{figure}[t]
\centering
%%% PLOT FILE - Group measures condition
\begin{tikzpicture}%
\begin{axis}[%
    ybar,
    %title=Test,%
    ylabel style={align=center},
    ylabel={Cooperation with robot \%},
    axis y line*=left, axis x line*=bottom,%
    ymin=0,
    ymax=1,
    ytick={0.2, 0.4, 0.6, 0.8, 1.0, 1.2},
    legend style={at={(1.50,1.1)}},
    %x tick label style={rotate=45,anchor=east},%
    symbolic x coords={Second.Player},%
    xtick=data,
    enlarge x limits=0.3,
    bar width=20pt,
    width=6.5cm,
    height=5.5cm,
    %ylabel=difference in \%,%
        ]%
        
\addplot+[%
    color=black, %
    fill={rgb,255:red,255; green,255; blue,255},%
    postaction={pattern= dots},
    error bars/.cd,%
    y dir=both,%
    y explicit,%
        ]%
coordinates {
            (Second.Player,0.81) +- (Second.Player,0.05)
          
        };%
\addplot+[
    color=black,
    fill={rgb,255:red,255; green,255; blue,255},%
    error bars, y dir=both, y explicit]
				coordinates {
			 (Second.Player,0.36) +- (Second.Player,0.06)
			};

\draw [signbarL] (axis cs:Second.Player,0.95) ++(.5*\pgfplotbarwidth, 0) -- +(-\pgfplotbarwidth, 0);
\draw (axis cs:Second.Player,0.95) node[above]{$\ast\ast$};

\legend{Cooperative,Non-cooperative}
\end{axis}%
\end{tikzpicture}%
\caption{Empirical results from the Child-Robot experiments. The bars represent the percentage of children that cooperated with the robot after observing a cooperative (left bar) or non-cooperative (right bar) action by the robot. We observe a statistically significant difference between the two conditions, suggesting that children are more likely to cooperate with a robot that was previously cooperative than with a robot that was non-cooperative. Error bars represent the standard error. ** $p<.001$\blue{.}}
\label{fig:cooperative}
\end{figure}

We used the independent-samples t-test to assess the differences between the two conditions. The results indicate a statistically significant difference in the mean cooperation rate of children playing with the robot after observing its cooperative or non-cooperative behaviour ($t = 5.457, p < .001$). Specifically, children who observed the robot behave cooperatively exhibited a significantly higher mean cooperation rate when playing with it afterward ($\hat{p} = 0.81$) compared to those who observed the robot being non-cooperative ($\hat{q} = 0.36$) (Figure \ref{fig:cooperative}). 

%The standard deviations for the two conditions are similar ($Cooperative\_SD= 0.40$ and $Non-Cooperative\_SD= 0.48$)

% detailed caption Children’s cooperation towards the robot depending on whether they had previously observed the robot behaving cooperatively or non-cooperatively with another child.

\subsection{Learning Model}

% Alexandre zone
The experimental results inform \blue{us} about the immediate impact of observing previous interactions \blue{on} a child\blue{'s} willingness to trust and cooperate with a robot. We used the $\hat{p}$ and $\hat{q}$ from the empirical study to inform the $p$ and $q$ of our models. Based on these data ($p=\hat{p}=0.81$, $q=\hat{q}=0.36$), we develop a model to infer the possibility that an artificial agent learns to trust and cooperate with humans in an indirect reciprocity setting. In particular, we focus on the capacity of different multi-armed bandit (MAB) learning algorithms to learn and maintain cooperative behaviour, based on a simulated scenario where the algorithm (representing the robot) plays a Stag Hunt game with an opponent that follows the previously observed behaviour of children (\blue{cooperate} or defect). \blue{Following the conducted experiment, opponents play once with the algorithm, following a single observation, in the limit of an infinite population, allowing us to discard strategic adaptations and memory effects.} We focus on one implementation of each of the three primary classes of MAB algorithms: $\epsilon$-greedy, UCB1, and Thompson Sampling (TS); see the Methods section for details on each of these models.

We start by performing hyperparameter tuning by selecting, for $\epsilon$-greedy and UCB1, the parameters that maximize cooperation for a mutual cooperation benefit of $b=2$ (see supplementary material). This corresponds to the scenario that mimics the experimental setup where two additional LEGO pieces are awarded following mutual cooperation. No additional tuning is performed throughout the experiments as a deliberate choice to evaluate the challenges of deploying algorithms trained on empirical data. %, extra benefit ($b=2$). %\textcolor{red}{For Thompson sampling, as its parameter $w$ corresponds to direct biases towards selecting a specific action and could therefore be used to always select cooperation, its priors are left at $w=0.5$, representing an equal initial probability to select either cooperation or defection.}

\begin{figure}[htb]\centering
    \includegraphics[width=0.7\columnwidth]{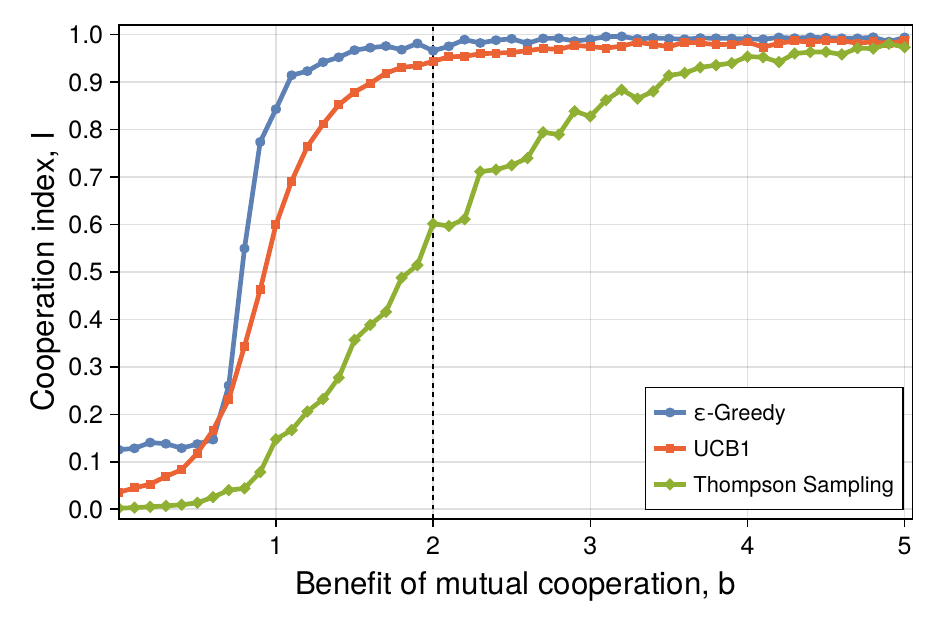} %width=0.83
    \caption{Simulation results: cooperation index, $I$ when varying the extra benefits of mutual  cooperation, $b$, for the three algorithms we utilize to simulate the robot learning to play with children. The dashed line, at $b=2$, indicates the benefit of mutual cooperation used in the experimental setting (where 2 additional LEGO pieces are given if mutual cooperation occurs). We observe that while cooperation is common and stable under $\epsilon$-greedy and UCB1 at a moderate $b > 1$, Thompson Sampling requires a greater benefit for cooperation to stabilize. Parameters used: $p =\hat{p} = 0.81$, $q =\hat{q} = 0.36$, $\epsilon = 1/128$, $c = 4$.}
\label{fig:bstudy}
\end{figure}

In Figure \ref{fig:bstudy}, we study the frequency of cooperation as we vary the extra benefits of mutual cooperation ($b$). This allows us to study the robustness of each algorithm in interaction contexts where coordination is easier (high $b$) or harder (low $b$). 
While $\epsilon$-greedy and UCB1 achieved high cooperation for $b = 2$, we observe that Thompson Sampling (TS) requires a substantially higher benefit of mutual cooperation for stable cooperation to emerge. Yet, at higher benefits, it achieves cooperation levels as high as those of the remaining algorithms. Furthermore, a lower benefit ($b<1$) leads to an near-zero level of cooperation across every algorithm tested.

%Having access to the fraction of cooperation of each algorithm, it is also possible to compute how frequently each algorithm received cooperation, allowing us to predict the action of children playing against a robot using a given algorithm. By definition, the frequency of cooperation received by the algorithm, $I^R$, is bounded by $min(p,q) \leq I^R \leq max(p,q)$, leading to a maximum value of $p$ ($q$) when always cooperating (defecting) having $p>q$ ($q>p$), and a minimum value of $q$ ($p$) when defecting (cooperating) instead. In practice, as $p>q$, this results in the cooperation received by the algorithm increasing in scenarios where $I>q$, and decreasing when $I>p$ (see supplementary material).

\begin{figure*}[tb!]\centering
\includegraphics[width=\textwidth]{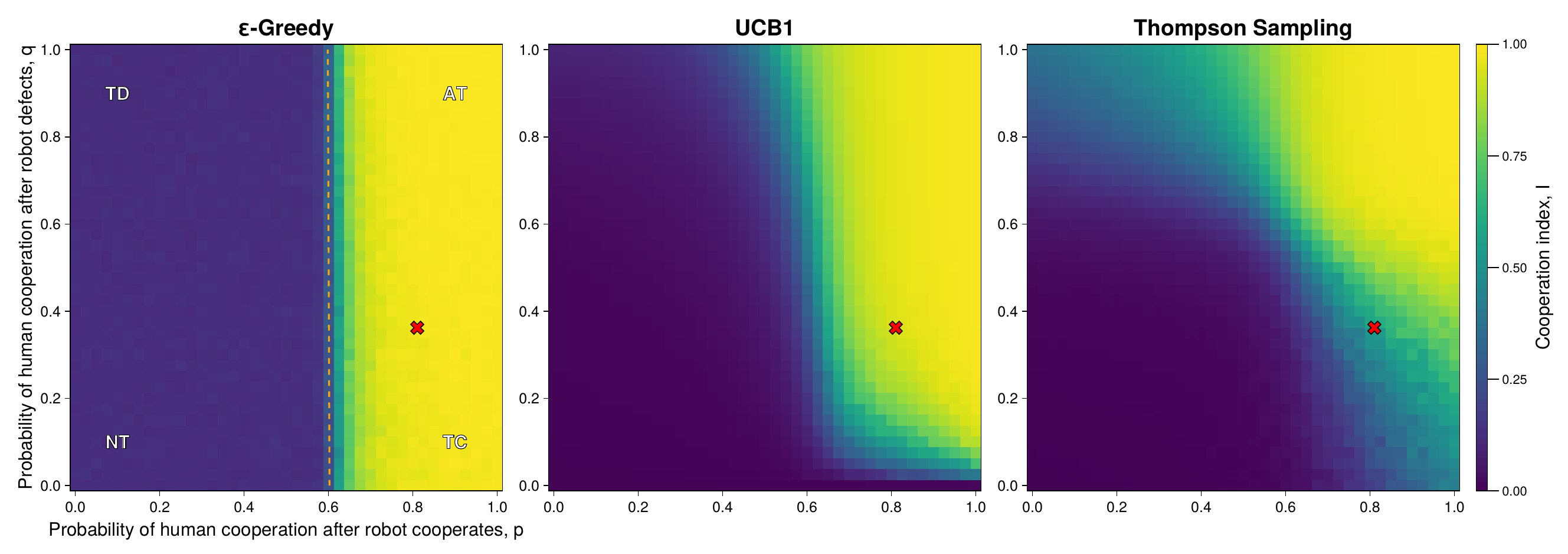} %width=0.83
\caption{Cooperation index, $I$, varying $p$ and $q$, for the three learning algorithms used. $p$ and $q$ represent, respectively, the probability that an agent cooperates with an opponent, after having observed the same opponent cooperating or defecting with a third-party. The red cross indicates the experimental value found ($p=\hat{p}=0.81$, $q=\hat{q}=0.36$). Labels indicate the pure strategy used by humans at that point: \textit{Always Trust, AT} ($p=1,q=1$), which always cooperates; \textit{Never Trust, NT} ($0,0$), which always defects; \textit{Trust Cooperators, TC} ($1,0$), which only cooperates if the algorithm previously cooperated; and \textit{Trust Defectors, TD} ($0, 1$), which only cooperates after observing a defection. The vertical line in $\epsilon$-greedy indicates its theoretical threshold for cooperation to dominate defection (see Methods). We observe that each algorithm has a different sensitivity to $p$ and $q$, resulting in distinct regions where each algorithm outperforms others. Parameters used: $b=2$, $\epsilon = 1/128$, $c = 4$.} 
\label{fig:pq_heatmap}
\end{figure*}

While the experimental results when facing a coordination dilemma, $\hat{p}$ and $\hat{q}$, provide an estimate for the current levels of reciprocity, $p$ and $q$\blue{,} respectively, it is also important to study alternative scenarios, corresponding to different cultural environments where these values can change. To this end, in Figure \ref{fig:pq_heatmap}, we set $b=2$ and instead vary $p$ and $q$. Crucially, we can assess how cooperation under each algorithm depends on $p$ and $q$: $\epsilon$-greedy is relatively independent of $q$ --- it is unaffected if children are more punishing following a defection from the algorithm ---, while displaying a hard threshold for high cooperation at around $p\geq0.60$ and otherwise showing low but non-zero levels of cooperation. As such, for $\epsilon$-greedy to learn to cooperate, it must be in an environment where children would choose to cooperate with the robot after observing it cooperate, with a probability $p\geq0.60$. UCB1 has a smoother decay of cooperation as $p$ goes down, with a higher resilience to low $p$ at high levels of $q$. Furthermore, UCB1 is sensitive to a low $q$, which can collapse cooperation if it approaches zero, highlighting a vulnerability against discriminating players that only cooperate after observing cooperation. This vulnerability arises from the method of exploration of the algorithm, which is impactful in early rounds: the high $c$ hyperparameter promotes exploration, which at a high enough value enables an always cooperative algorithm to explore defection, while, due to the payoff difference, not allowing a defecting algorithm to explore cooperation. 
Results under other parameterizations are presented in the supplementary material. Interestingly, Thompson Sampling (TS) presents a symmetry between $p$ and $q$, being able to sustain modest levels of cooperation at low $p$ \blue{and} high $q$ and vice versa. This means TS is capable of promoting cooperation even when only receiving cooperation following it playing defection. However, it is also equally sensitive to lower values of $q$, leading to lower cooperation when its defections are punished. As such, under TS, the region of very high cooperation is smaller than the remaining algorithms. 
These results highlight that no algorithm unconditionally outperforms all others in all circumstances.

Given the context in which the algorithms might be deployed, another aspect to consider is the time for an algorithm to converge. To this end, a comparison of the performance of each algorithm throughout the learning process is presented in the supplementary material. We show how not all algorithms learn equally fast, and how performance does not always improve in all regions as the number of rounds increases. In particular, $\epsilon$-greedy was the only algorithm to reduce a large region of cooperation (that is, the range of $p$ and $q$ that sustain cooperation) as the number of rounds increases, as it starts optimistically and then optimises its payoff.

% Isabel 
% fig 3 . resultados experiencia barra + valores 
\section{Discussion}\label{sec4}
In this section, we discuss \blue{whether} indirect reciprocity can be transposed to hybrid populations, and if robots can learn to cooperate using children's playing behaviour. Moreover, we reflect on the impact of our findings on robot design, and we describe the limitations of our work and future research avenues.

\subsection{Can indirect reciprocity be transposed to child-robot interactions? (RQ1)}
% Does indirect reciprocity apply to hybrid groups of robots and children? (RQ1); 
Our results indicate that children's trust and cooperative behaviour toward a robot are influenced by the robot's prior behaviour in similar situations — its reputation. This suggests that indirect reciprocity, as a cooperation mechanism, can extend to hybrid populations composed of artificial agents and children.
These findings align with previous literature demonstrating that children, from as early as 18–24 months old, use reputation and indirect reciprocity to guide their cooperation and decision-making \cite{Hamlin2011, jin2024infants, nava2019see}, even when the donors are artificial representations \cite{meristo2013infants}.

Another key finding was that children in the non-cooperative condition exhibited a cooperation rate of 0.36. This indicates that approximately one-third of these children were still willing to trust the robot, despite the high risk of losing their bricks, based on its reputation (stemming from its observation). They had previously seen the robot behave non-cooperatively, yet they still cooperate\blue{d} with it. This result is consistent with previous research on child-robot interaction, in which children chose to interact with a group of robots even after witnessing the same robots excluding another child \cite{correia24ostracism}.
Several factors may explain this outcome. First, robots often cause a novelty effect in children, which can influence their willingness to cooperate, a phenomenon well documented in the literature \cite{smedegaard2019reframing}. Second, the perceived value of rewards and risks may differ between children according to individual factors \cite{greening2005predictors} and their perceptions and knowledge about the machines \cite{vonshenk25Systematic}. For instance, some may find the risk of losing three bricks worthwhile in exchange for a potential gain of two, while others may adopt a more conservative approach and avoid any risk, in line with the reactions of adults to different payoff values and goals \cite{vonshenk25Systematic}.

\subsection{Can a robot learn to cooperate given the strategies revealed by children?  (RQ2)}

The gap between experimental and theoretical works in indirect reciprocity has recently received considerable attention. In particular, recent work has used both theoretical models to predict user studies \cite{santos2020picky}, and experimental results to inform theoretical models \cite{correia2025evolution, seinen2006social}. Yet, work related to human-robot interactions, and specifically with children, has not yet been discussed.

\blue{Based on} the one-shot environment used in the experimental scenario, we developed a repeated interaction theoretical multi-armed bandit (MAB) \cite{slivkins2019introduction} model that simulates a learning algorithm playing against an environment that follows the observed behaviour of children. 
In our model, the algorithm repeatedly plays a Stag Hunt game, where the probability \blue{of receiving} cooperation in a round is conditioned on its prior action. This allowed us to assess the capacity of an algorithm to learn cooperation in the long term under indirect reciprocity, as explored in our experiment. However, translating the experimental setup to the theoretical model required careful modelling choices both in the experiment and \blue{in} the model. In particular, the complexity of the experiment is constrained to measure data directly translatable to common features in indirect reciprocity models. In our work, these are the benefit of mutual cooperation and the strategy used by the children, which we then explore in the theoretical model. The difficulty for an algorithm to learn cooperation under such an environment lies in the relationship of the payoff of the current round with both the prior and current action of the algorithm, together with the non-deterministic action of the opponent. This can lead to scenarios where the action of the algorithm is incorrectly attributed a low or high payoff expectation as a consequence of the prior action.

Our results indicate that learning cooperation is not a guaranteed outcome, as it is dependent on all key variables of the model: the learning algorithm, the benefit of mutual cooperation versus that of defection, the frequency at which the algorithm receives cooperation after it cooperates or defects, and the hyperparameters of the algorithm. For the values used in the experimental user study, following a step of hyperparameter optimisation, we observe that $\epsilon$-greedy \cite{slivkins2019introduction} and Upper Confidence Bound (UCB1) \cite{auer2002finite} learn to cooperate with a high frequency ($I\approx0.95$ and $I\approx0.93$, respectively). Yet, Thompson Sampling (TS) \cite{thompson1933likelihood} achieves only a moderate level of cooperation ($I\approx0.60)$), highlighting the need for a careful selection of algorithms and hyperparameters. 

\subsection{What is the impact of different robotic behaviours and children's strategies on cooperation?  (RQ3)} 

%\todo{In general cooperation depends on the algorithm}
%\todo{We observe that some algorithms (e greedy and UCB1) can only learn cooperation if they have enough benefit. Yet TS is independent of it.}
%\todo{All algorithms have different sensitivity to the behaviour of humans. Some are more sensitive to how frequently they are rewarded following cooperation, others are sensitive to how frequently they are punished for not receiving coop}

Our findings reveal that different learning algorithms exhibit distinct sensitivities to the behaviour of children. Some algorithms primarily respond to how frequently they are rewarded following cooperation, as is the case with $\epsilon$-greedy, while others are highly influenced by the frequency with which they are punished for not cooperating, such as Thompson Sampling. 
We also study the learning process of each algorithm, showing that their behaviour throughout learning differs across all possible children strategies. While all algorithms generally improve cooperation over time, against some strategies, the algorithms might instead reduce cooperation throughout training as they adjust their expected payoff. As such, the learning rate and cooperation level of the algorithm are not only highly dependent on the algorithm itself, but also on the strategy of the children. This is particularly relevant as not all environments have frequent interactions, and as such the initial interactions with the algorithm might not reflect its future behaviour.

All this suggests that no optimal algorithm exists for all circumstances. As different interactions offer different benefits for mutual cooperation, and as the behaviour of children can change due to the normalisation or adaptation to technology \cite{smedegaard2019reframing}, or the cultural context in which they are inserted \cite{lim2021social}, a careful selection and parametrisation of the learning algorithm is necessary to ensure it is capable of promoting and maintaining prosocial behaviour. 

Finally, we are also capable of predicting the long-term behaviour of children throughout the learning process of the algorithm.
Since, when selecting their action, children consider the previous action of the robot, its strategy can influence the child’s willingness to engage. However, an algorithm that always cooperates does not guarantee cooperative responses from children, nor will an algorithm that always defects guarantee solely defective responses. This suggests a present but limited influence that robots can exert over human behaviour, aligning with previous experimental \cite{correia24ostracism, neto2023} and theoretical \cite{sharma2023small} research in other contexts.

\section{Limitations and Future Work}
%long-term effects
In this section, we acknowledge the limitations of our work.

\textbf{Long-term effects:} We measured children's cooperation immediately after observing another child's interaction with the robot. However, it remains unclear whether these effects persist over time with multiple games. Future studies should investigate whether prolonged exposure to a cooperative or non-cooperative robot leads to sustained changes in cooperative behaviour. 

\blue{\textbf{Cultural backgrounds of participants:}} We studied children from one school in one country. Although our findings may have broader applications, it is essential to consider the impact of cultural backgrounds on cooperation \cite{gachter2010culture, karpus2025human}. Future studies should explore different countries and settings for a more comprehensive understanding. 

\textbf{Empirical data as input to the models:} Using one-time empirical results as the foundation for evolutionary models is a novel approach for informing predictions. However, this method carries certain risks, as it assumes that a single observed outcome can be extrapolated to multiple interactions over time. Given that children’s responses—and robots' behaviours—may evolve with continued exposure, future work should consider longitudinal studies to refine these models. \blue{In addition, the behaviour of algorithms facing individuals who adapt their strategy following repeated interactions should also be considered.}

\textbf{Similar payoffs and rewards:} Our empirical model was based on a predefined payoff and reward system. In our study, the loss (3 bricks) was not substantially different from the potential gain (2 bricks). The observed cooperation rates might vary if the disparity between losses and gains were greater. Future research should explore how different payoff structures influence cooperative behaviour.

\blue{\textbf{Robot control:} In our empirical study, the robot was teleoperated by a researcher located outside the room, who selected the robot’s behaviours throughout the game. Although these behaviours were consistent — its voice was pre-recorded, and its movements were dependent on the condition — the researcher’s reaction time in triggering the robot’s responses could vary due to natural human delay, but this variation was not dependent on the experimental condition.}

\textbf{Social norm complexity:} In our experiment and model, we have considered that the only relevant factor for children to play the experimental game with the robot was the previously observed action done by the robot, not including other factors such as the action of the child playing with the robot, or any prior opinions that the observing child might have about the confederate child. This added complexity in the judgment rule used by an observer has a well-documented effect on the capacity to promote cooperation \cite{santos2018social} and should be considered in future work.

\textbf{Full observability and infinite populations:} The model we developed considers an environment where the prior action of the algorithm is always observed, and therefore always considered in the next interaction. Furthermore, our model does not capture possible effects from playing with an individual that was met before. However, in environments such as schools, not only are some actions only observed by parts of the population, leading to potential disagreements \cite{krellner2023we}, there are \blue{also} repeated interactions with the same individuals that might affect reputations \cite{santos_social_2016}. As it has been shown that finite populations and partial observability \cite{hilbe2018indirect, fujimoto2023evolutionary} deeply affect cooperative behaviour, future models should consider these variables.

\blue{\textbf{Explore other empirical scenarios:} Our work demonstrated the potential of using empirical data to inform theoretical models of cooperation. This novel approach raises questions about how it may apply to other scenarios. For example, how adults might respond to the same protocol, and whether robots modelling human behaviour and concerns can be transposed to adult populations. Additionally, models using other social dilemmas can also make use of empirical data.}

\blue{\textbf{Influence of trust and moral preferences:} Despite indirect reciprocity being triggered in children following observations \cite{meristo2013infants, nava2019see}, trust also plays a role in Stag Hunt games. Future experiments should study these effects in isolation, via other games such as the trust game \cite{kumar2020evolution}. To better understand the norms employed by children facing robots, studies of their moral preferences should include other aspects \cite{capraro2021mathematical}, such as truth-telling \cite{capraro2019evolution}.}

\section{Conclusion}\label{sec6}

Our study investigated indirect reciprocity in hybrid groups of children and robots. Our experimental findings demonstrate that children use reputation-based decisions when interacting with a robot, relying on their observations of the robot's prior cooperative behaviour with other children. Specifically, children were significantly more likely to cooperate ($\hat{p} = 81\%$) with a previously cooperative robot, whereas only $\hat{q} = 36\%$ of children chose to cooperate with a non-cooperative robot. These results align with previous research on the indirect reciprocity mechanism in children and its application to adults and artificial agents.

In addition, we explore a novel approach that utilizes empirical studies to inform learning dynamics models, highlighting both its potential and associated risks. This strategy could contribute to building more robust models; however, it also raises concerns about using one-time observations as definitive predictors for long-term interactions. Since both children and robots can adapt their cooperative behaviour over multiple sessions, relying solely on one-shot results may not accurately reflect long-term dynamics. Nevertheless, our model demonstrates that multiple algorithms can learn to cooperate, despite none of the tested algorithms outperforming the others across all parameter spaces. This highlights the need for a careful selection of algorithms, considering the context in which they are inserted.

Finally, we discuss the broader implications of our findings for researchers and practitioners developing decision-making algorithms for robots in child-centred environments, particularly regarding their long-term impact on cooperation.
\section{Methods}\label{sec5}

% Filipa - empirico
\subsection{Experimental Setup}
The experimental user study took place at a primary school in a controlled setting, using a single room of the school building specifically for the purpose of the experiment. Children participated individually as the goal was to test mechanisms of indirect reciprocity during a cooperative game between one child and a robot. \blue{We had 62 girls and 69 boys. Ten of the children were immigrants. The study took place in a public school in a Western European country, with participants representing a diverse socioeconomic background ranging from low to middle income levels.}

We used the well-known Stag Hunt game and re-framed it as a LEGO negotiation game to be more engaging for the children. In our adapted LEGO version of the Stag Hunt game, each player starts with three LEGO pieces and can either play all pieces or play no pieces, i.e., \textit{to cooperate / trust} or \textit{not cooperate / defect}, respectively. If both players cooperate, ten LEGO pieces were rewarded and divided between the players, which constitutes an increase from the initial amount of three to five LEGO pieces per player (an extra benefit of $b=2$). If both players defect, they keep their initial endowment of three LEGO pieces per player. If only one player cooperates, the defector keeps the initial endowment of three LEGO pieces and the cooperator loses all the pieces. As an incentive, children were told they would contribute to building a collective LEGO tower with the pieces they collected \blue{as a collective goal}.

To experimentally manipulate indirect reciprocity, each child first observed one demonstration game between the robot and a partner child, and only afterwards were they asked to play a new game themselves with the robot. While the partner child was asked to cooperate in the demonstration game, we manipulated the robot's action as the independent variable, which could be either to cooperate or to defect with the partner child. As a result, the study had two treatments/conditions in a between-subjects design, where each participant was either in the cooperative or the non-cooperative condition referring to the observed action of the robot when playing with another child before playing with them (i.e., the demonstration game).

The partner child of one experiment session was in reality the participant of the previous session. Because we collected children's participation sequentially within the same class and time slot, the first child of each class would not constitute a participant of our sample and only act as a partner.

The procedure involved two researchers: one was teleoperating the robot as a Wizard-Of-Oz and stayed outside the room, and the other was doing the briefing for the participants. In each session, which lasted approximately 10 minutes, the child was welcomed, met the robot, and then had the game rules explained. At this stage, children could ask questions and clarify the game rules with the researcher in the room. After being told the game instructions, the participant silently observed the demonstration game between the robot and the partner child (i.e., the previous participant) in one of the two conditions (i.e., cooperative and non-cooperative). The partner child placed their LEGO pieces in the collective tower (only if it was the cooperative condition), and was asked to leave the room. Then, the participant child would play the LEGO game with the robot and only the researcher was present in the room. In this game, the robot's action was always to cooperate, which means the participant would either get five LEGO pieces if they cooperated or three if they defected. After putting the LEGO pieces in the collective tower, the participant was asked to act as a partner for the next participant, and they were asked to play cooperatively in the demonstration game for the next child.

The chosen robot for this experiment was Dash\footnote{https://www.makewonder.com/pt/dash/} and its behaviours were scripted and predefined. Whenever the robot had to cooperate, it would verbally say ``I will play'' and move forward to pull the LEGO piece towards the centre. While defecting, the robot would verbally say ``I will not play'' and stay still in the same place.

Regarding the sample, a total of 131 children participated in the experiment. However, 15 children did not comply with the researcher's request in the demonstration game to play cooperatively (and instead chose to defect), which led us to discard 15 participants from the data analysis. All children had a written consent form signed by their legal guardians and were informed they could quit the activity at any time. Lastly, the experiment was approved by the Ethical Committee of our university.

% Alex - modelo 
\subsection{Learning Model}

% - We model the robot using multi-armed bandit. It can cooperate or defect. 
While the experimental setting allows us to assess how children choose to cooperate with a robot after observing its prior interaction, we now aim to assess under which learning algorithms and settings the robot can learn to cooperate. %We used the cooperation rate from the experimental settings, $\hat{p}$ and $\hat{q}$, and applied them in the theoretical model ($p=\hat{p}=0.81$, $q=\hat{q}=0.36$). 
In particular, we define a theoretical model using a multi-armed bandit (MAB) \cite{slivkins2019introduction} formulation similar to the experimental task, allowing us to study the frequency at which different algorithms opt to cooperate.  

% - this is the payoff table. This is how p and q work
To translate the experimental task to a MAB problem, in Table \ref{tab:payoff_matrix}, we present a generalized payoff table of the previously mentioned coordination game. We define $b$ to be the added benefit obtained when both agents cooperate in the game, with $b=2$ corresponding to the additional benefit used in the experiment. 

\begin{table}[h!]
\centering
\begin{tabular}{lllll}
                       &                       &                        & \multicolumn{2}{c}{Child}                              \\ \cline{4-5} 
                       &                       & \multicolumn{1}{l|}{}  & \multicolumn{1}{c|}{C}        & \multicolumn{1}{c|}{D}    \\ \cline{3-5} 
\multicolumn{2}{c|}{\multirow{2}{*}{Robot}} & \multicolumn{1}{c|}{C} & \multicolumn{1}{l|}{b+3, b+3} & \multicolumn{1}{l|}{0, 3} \\ \cline{3-5} 
\multicolumn{2}{c|}{}                          & \multicolumn{1}{c|}{D} & \multicolumn{1}{l|}{3, 0}     & \multicolumn{1}{l|}{3, 3} \\ \cline{3-5} 
                       &                       &                        &                               &                          
\end{tabular}
\caption{Payoff table for the experimental task. Any agent that defects keeps its three pieces. If an agent unilaterally cooperates, it loses its pieces. Otherwise if both agents cooperate, a benefit is added on top of the prior three pieces. In the experimental setup, we use $b = 2$.}
\label{tab:payoff_matrix}
\end{table}

The experimental setup contains two distinct phases: a first game where the robot and a partner child play while the participant observes, and the partner child always cooperates. The second game, the participant, the observer of the previous round, plays the game with the robot.
Since the fixed strategy of the partner child should not be considered during learning, as it is done for control purposes in the experiment, to model the learning of the robot we adopt a repeated version of this setup.
We consider a chained interaction structure where the algorithm (representing the robot) will repeatedly interact with another agent, while being observed by a third agent. In the next interaction, the previous round's observer plays against the agent, and a new observer enters the scene, which then plays in the following round.

As the agents have a probability $p$ and $q$ to cooperate following a cooperation and defection by the algorithm, respectively, we can consider a MAB with two actions: \textbf{C}, which has a probability $p$ ($q$) to reward $b+3$ if the previous action of the algorithm was \textbf{C} (\textbf{D}), rewarding $0$ otherwise; and \textbf{D}, which always rewards $3$. Our model defines a non-stationary distribution of payoffs for each action, resulting in a stochastic MAB \cite{besbes2014stochastic}. However, we adhere to algorithms that make no assumptions about the strategy of children, instead focusing on minimal algorithms with imperfect knowledge.

% - We use the 3 main approaches: e-greedy, UCB1 and thompson sampling. Explain implementation

We focus on 3 distinct algorithms: $\epsilon$-greedy \cite{slivkins2019introduction}, UCB1 \cite{auer2002finite}, and Thompson sampling (TS) \cite{thompson1933likelihood}. In $\epsilon$-greedy, a $Q$ vector is kept, which is continuously updated with the average reward obtained for each action. With a probability $(1-\epsilon)$, the action $a \in \{C,D\}$ with the highest $Q_a$ is picked, representing exploitation. Otherwise, with a probability $\epsilon$, a random action is picked, representing exploration. Given the simplicity of the algorithm, it is also possible to analytically derive when cooperation will payoff-dominate defection, which is given by $\left((1-\frac{\epsilon}{2})p + \frac{\epsilon}{2}q\right)(b+3) > 3$. 

UCB1 operates similarly to the greedy section of $\epsilon$-greedy, only instead picking the action that maximises $a_t = \arg\max_a \left( Q(a) + c \cdot \sqrt{\frac{\ln t}{N(a)}} \right)$, where $N(a)$ is the number of times action $a$ was played, $t$ is the total number of actions played, and $c$ is the exploration coefficient. 

Finally, TS employs a Bayesian approach where each action $a \in \{C, D\}$ is associated with a Beta distribution that represents the belief about the expected reward of that action. At each round, TS samples from these distributions to select the action with the highest sampled reward. We make use of the generalized stochastic implementation presented in \cite{agrawal2012analysis}, adapted by rescaling each received payoff $r$ to $\tilde{r} \in [0,1]$. Specifically, for each action, the algorithm samples a value $\theta_a$ from a $\text{Beta}(\alpha_a, \beta_a)$ distribution, where $\alpha_a$ and $\beta_a$ are parameters that encode the prior knowledge of the action’s success and failure counts, respectively. The action corresponding to the highest sampled value is then chosen. After each round, a Bernouli trial with a success probability $\tilde{r}$ is performed, and the action priors are updated based on the outcome: if the trial is successful $\alpha_a$ is increased by 1, otherwise, $\beta_a$ is increased by 1. While it is possible to initialize the priors using any distribution $w$, we initialize them all equally at $0.5$ to not enforce cooperation.

%Finally, TS employs a Bayesian approach where each action $a \in \{C, D\}$ is associated with a Beta distribution that represents the belief about the expected reward of that action. At each round, TS samples from these distributions to select the action with the highest sampled reward. Specifically, for each action, the algorithm samples a value $\theta_a$ from the Beta distribution $\text{Beta}(\alpha_a, \beta_a)$, where $\alpha_a$ and $\beta_a$ are parameters that encode the prior knowledge of the action’s success and failure counts, respectively. The action corresponding to the highest sampled value is then chosen. After each round, the action priors are updated based on the reward outcome: if the action is successful, which we consider when both agents cooperate or both defect, $\alpha_a$ is increased by 1, otherwise, $\beta_a$ is increased by 1. While possible to initialize the priors using any distribution $w$, we initialize them all equally at $0.5$ to not enforce cooperation.

% - We run N rounds for K simulations, and average
% - we measure cooperation as the number of actions C (maybe in the last X% of the actions?)

We run each algorithm for $2000$ rounds with $500$ simulations in each experiment. Experiments with more and less rounds are presented in the supplementary material. Afterward, an index of cooperation, $I$, is given by the fraction of times the algorithm picked action \textbf{C} throughout all rounds across all simulations. Similarly, we also define $I^R$ as the fraction of times that the algorithm received cooperation during the simulations, which represents the fraction of human cooperation when facing the robot.

% fig 1 .  detalhe da experiencia 

\section*{Acknowledgements}{We thank the schools and children involved in the study. We would like to thank the ELLIS Unit Amsterdam. This work was supported by FCT Research Units : Lasige UID/408/2025, Larsys  10.54499/LA/P/0083/2020 and 10.54499/UIDP/50009/2020, and INESC-ID UIDB/50021/2020. F.P.S. acknowledges funding through the project "Reputation as a new route to cooperation in multi-agent reinforcement learning" with file number OCENW.M.22.322 of the research programme Open Competition Domain Science which is (partly) financed by the Dutch Research Council (NWO)). F.C acknowledge funding by FCT 2022.00816.CEECIND/CP1713/CT0013. I.N acknowledge funding by European Tailor Project (GA 952215).} 

%%%%%%%%%% Insert bibliography here %%%%%%%%%%%%%%

\bibliographystyle{plain}
\bibliography{reference}

\end{document}